\documentclass[amssymb,aps,prd,superscriptaddress,nofootinbib,twocolumn,preprintnumbers]{revtex4-1}
\pdfoutput=1
\usepackage[utf8]{inputenc}
\usepackage{amsmath}
\usepackage{amssymb}
\usepackage{xcolor}
\usepackage{graphicx, epsfig, amssymb} 
\usepackage{moresize}
\usepackage{hyperref}

\def\be{\begin{equation}}
\def\ee{\end{equation}}

\begin{document}
\preprint{CERN-TH-2016-257, INR-TH-2016-049}
\title{Ultra-Light Dark Matter Resonates with Binary Pulsars}
   \author{Diego Blas}\email{Diego.Blas@cern.ch}
 \affiliation{Theoretical Physics Department, 
CERN, CH-1211 Gen\`eve 23, Switzerland}

   \author{Diana L\'opez Nacir}\email{Diana.Laura.Lopez.Nacir@cern.ch}
 \affiliation{Theoretical Physics Department, 
CERN, CH-1211 Gen\`eve 23, Switzerland}

   \author{Sergey Sibiryakov}\email{Sergey.Sibiryakov@cern.ch}
 \affiliation{Theoretical Physics Department, 
CERN, CH-1211 Gen\`eve 23, Switzerland}
 \affiliation{Institute of Physics, LPPC, Ecole Polytechnique 
F\'ed\'erale de Lausanne, CH-1015, Lausanne, Switzerland}
 \affiliation{Institute for Nuclear Research of the Russian Academy
of Sciences, 60th October Anniversary Prospect, 7a, 117312  Moscow, Russia}

\begin{abstract}
We consider the scenario where dark matter (DM) is represented by an
ultra-light classical scalar field performing coherent periodic
oscillations. We point out that such DM perturbs the dynamics of
binary systems either through its gravitational field or via direct
coupling to ordinary matter. This perturbation gets resonantly amplified if
the frequency of DM oscillations is close to a (half-)integer multiple
of the orbital frequency of the system and leads to a secular variation of the
orbital period. We suggest to use binary pulsars as probes of this
scenario and estimate their sensitivity. While the current accuracy of  observations
is not yet sufficient to probe the purely gravitational
effect of DM, it already yields  constraints on direct
coupling that are competitive with other bounds. The sensitivity
will increase with the upcoming radio observatories such as Square
Kilometer Array. 
\end{abstract}

\maketitle

\paragraph*{{\bf Introduction} -}

Despite years of intensive research, the nature of
dark matter (DM) remains unknown. An interesting possibility is that DM
is represented by a very light boson with extremely
weak, if any, coupling to the fields of the Standard Model,
see~\cite{Marsh:2015xka} for recent review. Such DM candidates are
common in many models of new physics including the axion solution to the
strong CP problem \cite{Peccei:1977hh,Wilczek:1977pj,Weinberg:1977ma}, 
the relaxion mechanism for the
origin of the electroweak symmetry breaking \cite{Graham:2015cka}, 
and string theory \cite{Svrcek:2006yi,Arvanitaki:2009fg}. 
Huge particle occupation
numbers required to reproduce the DM density imply that 
such ultra-light dark matter (ULDM) is 
well described by a classical scalar field $\Phi$. 
For a single field to be all of the DM,  the  anharmonicities  of the
potential are constrained to  be small when the field starts
oscillating at $H\sim m_{\Phi}$ (with $H$ the Hubble rate) 
\cite{Arvanitaki:2014faa}.  
Since the amplitude of the field at that moment is larger than its
present amplitude in the galactic halo, we will neglect possible
self-interactions of~$\Phi$. 

A lot of effort has been devoted to identify observations
and experiments sensitive to ULDM. 
Very light candidates with masses $m_\Phi\lesssim
10^{-24}$\,eV are excluded as the dominant DM component by the
observations of the cosmic microwave background and large-scale
structure~\cite{Hlozek:2014lca};  future observation are expected to push the lower bound to  
$m_\Phi\sim 10^{-23}$\,eV \cite{Hlozek:2016lzm}.
Slightly heavier ULDM in the range $m_\Phi\sim 10^{-23}\div 10^{-21}$\,eV can be probed by the
Lyman-$\alpha$ forest, galaxy formation history and  
the structure of galactic halos 
\cite{Hu:2000ke,Amendola:2005ad,Bozek:2014uqa,Schive:2015kza,Sarkar:2015dib,Hui:2016ltb};
a complementary probe is
provided by Pulsar Timing Arrays (PTA) 
\cite{Khmelnitsky:2013lxt,Porayko:2014rfa}. 
The mass range up to 
$m_\Phi\sim 10^{-18}$\,eV can, in principle, be accessible to 21cm surveys \cite{Marsh:2015daa}. 
Scalar
fields with $m_\Phi\sim 10^{-14}\div 10^{-10}$\,eV would be produced
by rotating stellar mass black holes via superradiance which implies
various observable signatures including gravitational wave emission in
the LIGO sensitivity band \cite{Arvanitaki:2010sy,Arvanitaki:2016qwi}.
Future studies of supermassive black holes can potentially access
lighter masses $m_\Phi\sim 10^{-20}\div 10^{-15}$\,eV.
The 
gravitational effect of ULDM on laser interferometers was explored in~\cite{Aoki:2016kwl}.

The previous observations probe purely gravitational interactions of ULDM. If
ULDM has a direct coupling to ordinary matter the
possibilities to test it are more diverse and depend on specific
models.
A rather generic effect of ULDM is periodic modulation of
the Standard Model couplings and particle masses with time. A number
of proposals have been recently put forward to search for such variations
using atomic
clocks
\cite{Derevianko:2013oaa,Arvanitaki:2014faa,VanTilburg:2015oza,Stadnik:2015kia,Hees:2016gop,Stadnik:2016zkf},
accelerometers \cite{Graham:2015ifn}, resonant-mass detectors 
\cite{Arvanitaki:2015iga}, laser and 
atom interferometry
\cite{Stadnik:2014tta,Stadnik:2015xbn,Geraci:2016fva,Arvanitaki:2016fyj}.

In this work we propose to use observations of binary pulsars as a probe of
ULDM in the mass range $m_\Phi\sim 10^{-23}\div 10^{-18}$\,eV. The
exquisite precision of the measurements combined with the clean
theoretical description makes of binary pulsars highly sensitive to new
physics that affects dynamics of massive objects
\cite{Manchester:2015mda,Kramer:2016kwa}. This property has been
already exploited to constrain alternatives to general relativity  
\cite{Wex:2014nva,Will:2014xja} and led to the suggestion to use binary pulsars as
resonant detectors of the stochastic gravitational wave (GW) background 
\cite{Hui:2012yp} (see \cite{Bertotti1973,1975SvA....19..270R,
1978ApJ...223..285M,Turner:1979yn,Mashhoon:1981wn} for related earlier
works). The influence of DM composed of
  weakly interacting 
heavy particles on  the dynamics of binary pulsars was studied in \cite{Pani:2015qhr}.

The main idea of our approach is close in spirit to 
Refs.~\cite{Hui:2012yp,Bertotti1973,1975SvA....19..270R,1978ApJ...223..285M,Turner:1979yn,Mashhoon:1981wn} 
and can be summarized as follows. The ULDM field $\Phi$ in
the galactic halo
represents a collection of plane waves with frequencies\footnote{We
  use units $c=\hbar=1$.}
$\omega_\Phi\simeq m_\Phi+m_\Phi v^2/2$ and momenta $k_\Phi\simeq
m_\Phi v$, where $v\sim 10^{-3}$ is the typical virial velocity in the
halo. Neglecting the term $m_\Phi v^2/2$ in the
frequency we obtain the general form of the ULDM field,
\be
\label{eq:phidm}
\Phi({\bf x},t)=\Phi_0({\bf x})\cos\big(m_\Phi t+\Upsilon({\bf
  x})\big)\;,
\ee
where $\Phi_0({\bf x})$ and $\Upsilon({\bf x})$ are slowly varying
functions of position. A binary system embedded in the DM background
(\ref{eq:phidm}) will experience periodic perturbation due to the
change in the gravitational field of $\Phi$ and, in the presence of a
direct coupling, due to the change in the masses of the stars in the
binary. 
If the frequency of the perturbation happens to be close to an integer
multiple of the binary orbital frequency, its effect is resonantly
amplified  
and leads to a secular change in the orbital period 
that can be searched for experimentally. 
 We now proceed to the quantitative discussion. We
start with the case when DM and ordinary matter interact only
gravitationally.

\paragraph*{\bf ULDM interacting only through gravity -}
The energy-momentum of a free massive oscillating field
(\ref{eq:phidm}) corresponds to the density and pressure \cite{Khmelnitsky:2013lxt},
\be
\label{rhopDM}
\rho_{DM}=\frac{m_\Phi^2\Phi_0^2}{2},\qquad
p_{DM} =-\rho_{DM}\cos(2m_\Phi t+2\Upsilon)\;. 
\ee 
The latter generates an oscillating perturbation of the metric. To
find this
we use the Newtonian gauge,
\be
\label{eq:FRW}
ds^2=-(1+2\phi) dt^2+(1-2\psi)\delta_{ij}dx^idx^j\;,
\ee
and write down the trace of the $(ij)$ Einstein equations,
\[
6\ddot\psi+2\Delta(\phi-\psi)=24\pi G p_{DM}\;.
\]
Neglecting the spatial gradients and using (\ref{rhopDM}) we obtain,
\begin{equation}
\ddot{\psi}=-4\pi G\rho_{DM}\cos(2m_\Phi t+2\Upsilon)\;. 
\label{eq:ddpsi}
\end{equation}  
This can be viewed as a standing scalar GW. Similarly
to the usual GW's, it produces an extra relative acceleration between
the bodies in a binary system. This is conveniently written in the
Fermi normal coordinates associated to the center of mass of the
binary~\cite{1978ApJ...223..285M}, 
\be
\label{gravaccel}
\delta \ddot r^i=-\delta R^i_{\phantom{i}0j0}r^j=-\ddot\psi\, r^i\;,
\ee
where $r^i$ is the vector connecting the two bodies and 
$\delta R^i_{\phantom{i}0j0}$ is the contribution of GW into the
corresponding components of the Riemann tensor. In the last equality
we  evaluated $\delta R^i_{\phantom{i}0j0}$
in the conformal gauge~(\ref{eq:FRW})  since it is coordinate independent  at the linearized level.

Next, we compute 
the change in the energy of a binary system with masses $M_{1,2}$  during one orbital period $P_b$ due to its interaction with
ULDM,
\[
\begin{split}
\delta E_b&=\mu\int_0^{P_b}\dot r^i\delta\ddot r^idt\\
&=4\pi G\rho_{DM}\mu\int_0^{P_b} \dot r(t)r(t)
\cos(2m_\Phi t+2\Upsilon)dt\;,
\end{split}
\]
where $r$ is the distance between the bodies and
$\mu\equiv\frac{M_1M_2}{M_1+M_2}$ is the reduced mass of the
system. The energy exchange is most efficient when the orbital period
is close to an integer multiple of the period of metric
oscillations. Given that $P_b\propto |E_b|^{-3/2}$, the change in Keplerian energy leads to a
secular drift of the orbital period. Defining
\be
\label{deltaomega}
\delta\omega=2m_\Phi-2\pi N/P_b~,~~~~
|\delta\omega|\ll 2m_\Phi\;,
\ee
and using the standard formulas of Keplerian mechanics we obtain
the time derivative of the period averaged over time intervals $\Delta t$ satisfying
$P_b\ll\Delta t\ll 2\pi/\delta\omega$,
\begin{align}
\label{dotwgrav2} 
\langle \dot{P_b}& \rangle=-6G\rho_{DM}P_b^2\,\frac{J_N(Ne)}{N}f(t)\\
&\simeq -1.6\times 10^{-17}\left(\frac{ \rho_{DM} }{0.3\, \frac{{\rm
        GeV}}{\rm cm^3}}\right)
\left(\frac{P_b}{100\,{\rm d}}\right)^2  \frac{J_N(Ne)}{N}f(t)
\,,\notag
\end{align}
where
\[
f(t)=\sin\big(\delta\omega\, t+2m_\Phi t_0+2\Upsilon\big)\;,
\]
$J_N(x)$ are Bessel functions, $e$ is the orbital eccentricity, and
$t_0$ is the time of the first periastron passage 
since $t=0$. In the second line of (\ref{dotwgrav2}) we have normalized
$\rho_{DM}$ to the local DM density $\sim 0.3\,{\rm GeV/cm}^3$
in the neighborhood of the Solar System. 
We observe that, depending on the relative phase
between the orbital motion and the ULDM oscillations, the sign of $\langle \dot
P_b\rangle$ can be negative (decrease  of the binary system
energy) or positive (increase of the energy). Furthermore, the sign
alternates in time with the period $2\pi/\delta\omega\gg P_b$ which can be
used to discriminate this effect from other contributions to the
measured $\dot P_b$, such as e.g. those due to
the acceleration of the binary with
respect to the Solar System. 

\begin{figure}[h]
 \includegraphics[width=0.47\textwidth]{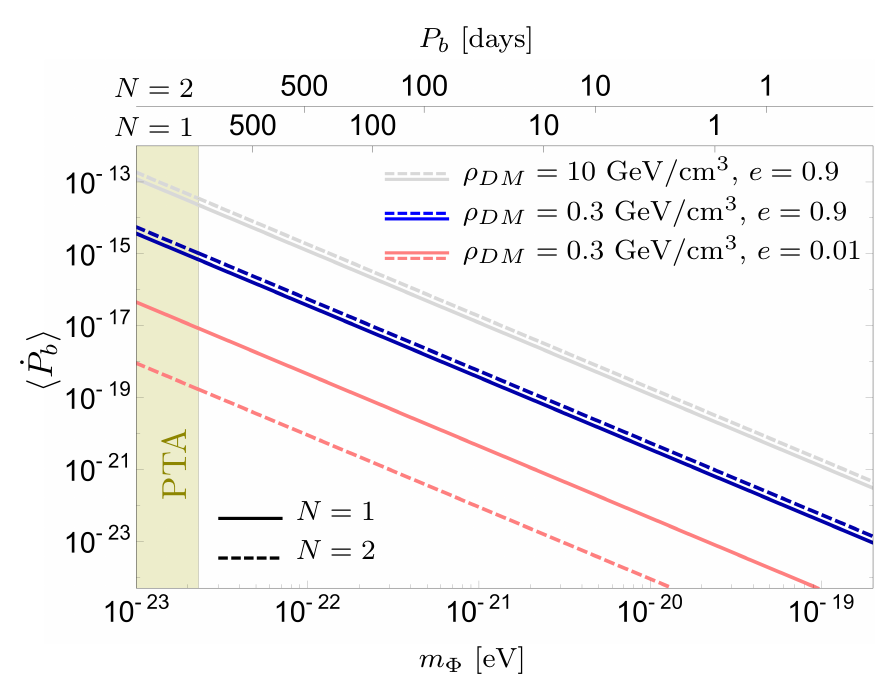} 
 \caption{\label{fig:consG}
Secular derivative of the orbital period given in
eq.~\eqref{dotwgrav2} 
as a function of the dark matter mass. 
We have set $f(t)=-1$ for the
numerical estimate.
Solid lines  assume resonances for $N=1$ ($m_{\Phi}=\pi/P_b$), while
dashed ones are for $N=2$ ($m_{\Phi}=2\pi/P_b$). The
corresponding orbital periods are shown on the two top axes. 
  The pink (lower) lines correspond to $\rho_{DM}=0.3
  \,\mbox{GeV}/\mbox{cm}^3$ and $e=0.01$, the blue  (middle)  lines are for the
  same $\rho_{DM}$ but $e=0.9$, while the grey (upper) lines correspond
  to $\rho_{DM}= 10\, \mbox{GeV}/\mbox{cm}^3$ and $e=0.9$.  The olive
  band on the left marks the regions $m_{\Phi}\lesssim 2.3 \times 10^{-23}$eV that
  can be probed by future pulsar timing
  arrays \cite{Khmelnitsky:2013lxt}. 
}
\end{figure}

The expression (\ref{dotwgrav2}) implies that the effect vanishes for
circular orbits ($e=0$) and grows with the orbital
eccentricity. Besides, it is stronger for systems with large orbital
periods. These points are illustrated in Fig.~\ref{fig:consG}.
We see that slow non-relativistic systems with orbital periods of tens
to hundreds of days and high eccentricity present suitable targets to
search for ULDM in the mass range $m_\Phi=10^{-23}\div 10^{-21}$eV. At
present there is a dozen of known binary pulsars satisfying these
requirements \cite{pscat,*2005AJ....129.1993M};
this number is expected to increase
dramatically with the operation of the Square Kilometer Array
\cite{Kramer:2015bea}. Note that for such systems the strength of the resonance
on higher harmonics ($N\geq 2$) is comparable to the strength of the
main resonance ($N=1$), which implies that a single eccentric binary
probes several different ULDM masses.

Detecting $\langle \dot P_b\rangle$ induced by pure gravitational
interaction of ULDM will be challenging. It will require the accuracy
of at least $10^{-16}$ in the determination of this quantity for
non-relativistic binaries. Currently such  precision has
been achieved for the double pulsar PSR J0737-3039A/B 
\cite{Kehl:2016mgp}, whose orbital period is, however, too short 
($P_b\approx 0.1$\,d) to be sensitive to the gravitational effect of
ULDM (the situation is different in the presence of  direct coupling,
see below). One may hope that the peculiar periodic modulation of
$\langle \dot P_b\rangle$  predicted by (\ref{dotwgrav2}) 
and the
expected correlation of its phase among systems located within the
coherence length of ULDM can be used to increase the sensitivity of
the search. 
A proper estimation of the measurability of these effects is beyond
the purpose of this {\it letter}. Still, let us comment on what could be an optimal scenario. For the modulation to be in principle detectable one
needs to observe the system for at least half of the period $T_{\rm
  obs}\gtrsim T_{\rm mod}/2=\pi/\delta\omega$.  The uncertainty in the measurement of $\langle \dot P_b\rangle$ for a constant secular drift scales as $T_{\rm obs}^{-5/2}$  \cite{Damour:1991rd}. For a modulated
signal, the drift in $P_b$ can be considered as approximately constant for half of the period. So, we will take as a conservative estimate for the error $\Delta\langle \dot P_b\rangle\propto (T_{\rm mod}/2)^{-5/2}$.  Thus, long modulations are preferable for
detection (the closer the system is to the resonance, the better);
still, they should stay within the range 
$T_{\rm mod}\lesssim 2 T_{\rm obs}$. An additional handle is provided by ULDM-induced secular variations in other orbital parameters \cite{longpap}.

A complementary way to improve the sensitivity to ULDM is to look for binary
systems in a denser DM environment. The  Navarro--Frenk--White halo profile 
\cite{Navarro:1995iw}
 predicts
an increase of $\rho_{DM}$ up to $\sim 10\,{\rm GeV/cm}^3$ within
$0.5$\,kpc distance from the Galactic Center\footnote{In the ULDM
  scenario the inner part of the Milky Way halo can contain a
  solitonic core which may further increase $\rho_{DM}$ in the
  vicinity of the
Galactic Center \cite{Schive:2014dra}. However, the size of the core
is smaller than $0.5$\,kpc.}. The impact of such 
increase on $\langle \dot P_b\rangle$ is shown in 
Fig.~\ref{fig:consG} by grey lines.

The possibility to discover the previous effects in future measurements 
strongly depends on the characteristics of
each individual observed system. Since these are currently unknown, a precise
forecast of detectability is impossible  at the present stage. An insight can be gained
from the analysis of simulated mock samples. This study must be
performed in the future to assess the actual measurability of the purely
gravitational effect.

\paragraph*{\bf ULDM directly coupled to matter -}

We now assume that ULDM interacts directly with the bodies in the binary
by affecting their masses,
\be
\label{masscoupl}
M_{1,2}(\Phi)=M_{1,2}\big(1+\alpha(\Phi)\big)~,~~~~~
|\alpha(\Phi)|\ll 1\;.
\ee
For simplicity, we focus on the case of universal  coupling\footnote{Namely, we assume all particle species couple to the same effective metric that depends on the scalar field and hence the weak equivalence principle is preserved.}, the
 case of different couplings will be treated elsewhere
\cite{longpap}. In what follows we will neglect the gravitational
interaction between ULDM and the binary. Then in the non-relativistic
limit the system is described by the Lagrangian,
\[
L=M_1(\Phi)\bigg(-1+\frac{v^2_1}{2}\bigg)
+M_2(\Phi)\bigg(-1+\frac{v^2_2}{2}\bigg)
+\frac{G M_1(\Phi) M_2(\Phi)}{r}.
\]
By combining  the equations of motion of the two bodies we obtain that
their relative acceleration acquires a contribution proportional to
the direct coupling,
\begin{equation}
\label{eq:r12}
\delta\ddot r^i=-\frac{d\alpha}{d\Phi}\,\dot\Phi\,\dot r^i
-\alpha(\Phi)\frac{G(M_1+M_2)r^i}{r^3}\;.
\end{equation}
As in the case of the pure gravitational interaction, this leads to the
change in the Keplerian energy and hence a secular drift of the
orbital period. Below we consider two choices for the function
$\alpha(\Phi)$. 

\paragraph*{Linear coupling $\alpha(\Phi)=\Phi/\Lambda_1$.}
In this case the condition for the resonance reads,
\[
\delta\omega=m_\Phi-2\pi N/P_b~,~~~~~
|\delta \omega|\ll m_\Phi\;.
\]
Evaluating the energy change due to (\ref{eq:r12}) and relating it to
the derivative of the orbital period we obtain,
\be
\label{eq:dotalpha}
\begin{split}
\langle\dot{P_b}\rangle &\simeq \, 2.5\times10^{-12}  
\bigg(\frac{\rho_{DM}}{{0.3\, \frac{{\rm GeV}}{\rm cm^3}}}
\bigg)^{\frac{1}{2}}
\left(\frac{P_b}{100\, {\rm d}}\right)\\
&\times\bigg(\frac{10^{23} \mathrm{GeV}}{\Lambda_1}\bigg)
J_N(Ne)\sin(\delta\omega\,t+m_\Phi t_0+\Upsilon).
\end{split}
\ee
\paragraph*{Quadratic coupling $\alpha(\Phi)=\Phi^2/(2\Lambda_2^2)$.}
Here we are back to the resonant condition (\ref{deltaomega}) and the
ULDM-induced variation of the orbital period is,
\be
\label{eq:dotquadr}
\begin{split}
&\langle\dot{P_b}\rangle \simeq \, 1.1\times10^{-11}  
\bigg(\frac{\rho_{DM}}{{0.3\, \frac{{\rm GeV}}{\rm cm^3}}}
\bigg)
\left(\frac{P_b}{100\, {\rm d}}\right)^2\\
&~\times\!\bigg(\frac{10^{16} \mathrm{GeV}}{\Lambda_2}\bigg)^2
\frac{J_N(Ne)}{N}\sin(\delta\omega\,t+2m_\Phi t_0+2\Upsilon).
\end{split}
\ee
Similar to eq.~(\ref{dotwgrav2}), 
the expressions (\ref{eq:dotalpha}) and (\ref{eq:dotquadr}) 
vanish for circular orbits implying that
systems with higher eccentricity are preferred to search for the
effect.

Current constraints on $\Lambda_{1,2}$ come from several sources.
Linear coupling to a light scalar field modifies the
attraction between massive bodies. This occurs even if $\Phi$ is not the 
DM and has been constrained by Doppler tracking of the Cassini
spacecraft \cite{Bertotti:2003rm} yielding 
$\Lambda_1\gtrsim 10^{21}$\,GeV. On the other hand, this bound does
not directly apply to the quadratic coupling\footnote{\label{footCa}The presence of a non-zero background of the $\Phi$-field within the Solar System would lead to a generation of a linear time-dependent $\Phi$-coupling. In principle, this effect can be used to put constraints on $\Lambda_2$ from a reanalysis of the Cassini data of \cite{Bertotti:2003rm}.} leaving a much milder constraint
from astrophysical processes and short-distance tests of gravity
$\Lambda_2\gtrsim 10^{4}$\,GeV 
\cite{Olive:2007aj,Derevianko:2013oaa}.
For a scalar field comprising ULDM
additional bounds arise due to the constraints on the GW
background. 
Indeed, the direct coupling of the scalar field to
masses (\ref{masscoupl}) can be absorbed by a redefinition of the
metric to the so-called Jordan frame
\[
g_{\mu\nu}\mapsto \bar g_{\mu\nu}=g_{\mu\nu}\big(1+2\alpha(\Phi)\big)\;.
\]  
\begin{figure}[h!] 
\includegraphics[width=0.47\textwidth]{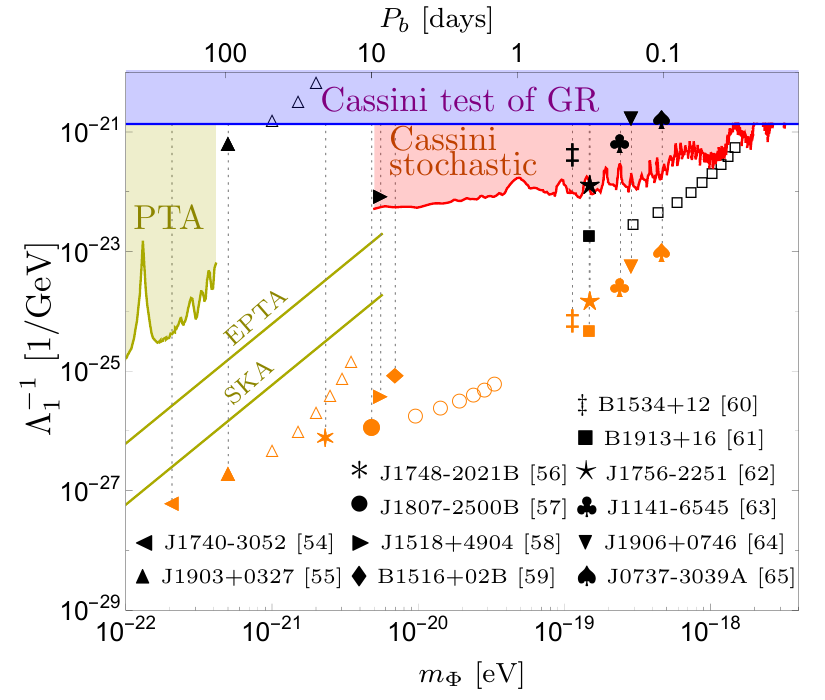} 
 \caption{\label{fig:consL} Sensitivity of binary pulsar observations
   to the linear coupling $\Lambda_1^{-1}$ between ULDM and ordinary
   matter for several known systems (see the indicated
   references for their description). 
Black symbols are constraints derived using the existing data on
$\dot P_b$, from 
\cite{Madsen:2012rs,Freire:2010tf,Freire:2007jd,Lynch:2011aa,Janssen:2008mh,Freire:2007xg,Fonseca:2014qla,Weisberg:2016jye,Ferdman:2014rna,Bhat:2008ck,vanLeeuwen:2014sca,Kramer:2006nb}; values above the symbols
are excluded. Orange symbols show the sensitivity that can be
achieved assuming $\dot P_b$ is measured for a given system with the accuracy $10^{-16}$ (see comments on the feasibility of such measurements in the main text). Empty symbols 
correspond
to resonances on higher harmonics ($N\geq 2$).   
The colored regions of the ULDM parameter space are excluded by PTA
\cite{Porayko:2014rfa} (olive), Cassini test of general relativity
\cite{Bertotti:2003rm} (violet) and Cassini bound on stochastic GW background
\cite{2003ApJ...599..806A} (red). 
Olive lines show future sensitivities of European Pulsar Timing Array
(upper) 
and Square Kilometer Array (lower) as
estimated in \cite{Graham:2015ifn}.}
\end{figure}

\begin{figure}[t!]
 \includegraphics[width=0.47\textwidth]{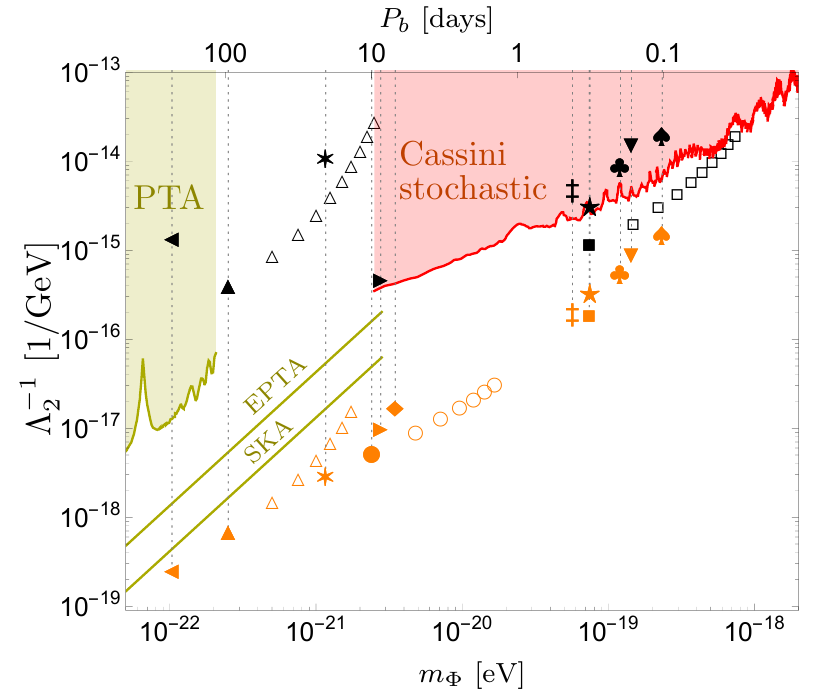}
 \caption{\label{fig:cons2} Same as Fig.~\ref{fig:consL}, but for the case
   of quadratically coupled ULDM. Recall that the possible constraints on
   $\Lambda_2^{-1}$ from Solar System tests have not been derived yet (see footnote \ref{footCa}).}
\end{figure}
Test bodies move along geodesics of $\bar g_{\mu\nu}$ which has an
oscillatory component due to oscillations of $\Phi$ --- this is a
scalar GW. The amplitude of such oscillations
has been constrained
by PTA data \cite{Porayko:2014rfa} in the low frequency region
($f<10^{-7}$\,Hz) 
 and by Cassini tracking\footnote{Ref.~\cite{2003ApJ...599..806A} 
presents the limits on the amplitude $h_c$ of
the stochastic background of { transverse} GW's that can be translated
 into a bound on the  scalar GW amplitude $\psi_c$ by
identifying $h_c\simeq\sqrt{15}\psi_c$.
This is slightly different from the
  identification $h_c\simeq 2\sqrt{3}\psi_c$ used in
  \cite{Khmelnitsky:2013lxt,Porayko:2014rfa} because, unlike pulsar
  timing, the light-time between the 
 spacecraft and the Earth is less than the GW
  period, which leads to a different expression for
  the average of the stochastic signal.}  
\cite{2003ApJ...599..806A}
in the 
frequency range $f=10^{-6}\div 10^{-3}$\,Hz.

Limits on an ULDM-induced contribution into $\dot P_b$ in the timing
model of binary systems can be used to put further bounds on the
couplings $\Lambda_{1,2}^{-1}$. Taking the reported error in the determination
of intrinsic $\dot P_b$ for several known systems as an upper limit on the
ULDM-induced contribution we obtain the constraints presented in
Figs.~\ref{fig:consL},\,\ref{fig:cons2}. 
In deriving them we have set the oscillating
factors in (\ref{eq:dotalpha}), (\ref{eq:dotquadr}) 
to one for the sake of the argument.  
One observes that they are competitive with the existing bounds.
In particular, the Hulse--Taylor pulsar
B1913+16~\cite{Weisberg:2016jye} provides the most sensitive probe of
the direct ULDM coupling for $m_\Phi$ satisfying the appropriate
resonance condition; whereas the systems 
J1903+0327~\cite{Freire:2010tf} and J1748-2021B~\cite{Freire:2007jd}
give the strongest constraints on the quadratic coupling 
$\Lambda_2^{-1}$ in the
range of $m_\Phi$ from $2\times 10^{-22}$\,eV to 
$2\times 10^{-21}$\,eV.
The
situation will further improve with the increase of precision in binary pulsar
timing and discovery of new binary systems. To illustrate this, we report the bounds one would obtain assuming that the systems considered before are timed at the best current precision (orange symbols in Figs.~\ref{fig:consL}  and \ref{fig:cons2}). Realistically, this precision may not be achievable for some of the  systems presented in the plots due to various sources of intrinsic uncertainty in determination of $\dot{P}_b$. Nevertheless, it is reasonable to expect that future surveys will discover new binary pulsars suitable for precision timing and having periods and eccentricity similar to those of already existing systems.   
 Note that valuable constraints on ULDM with masses 
$m_\Phi\sim 10^{-19}\div 10^{-18}$\,eV  can be obtained even from 
fast binaries
with periods down to a few hours.
In the low-mass region binary pulsar observations can be
complementary to future PTA.

The precise values of the bounds from Figs.~\ref{fig:consL},\,\ref{fig:cons2}  
should
be taken with caution. First, we have set the sine factors in
(\ref{eq:dotalpha}), (\ref{eq:dotquadr}) to one, whereas their
accidental suppression 
if the phase happens to be close to an integer multiple of $\pi$ is
not excluded. This option would be reliably ruled out by studying an
ensemble of systems, which would allow to average over the phases. At the moment such study is impossible due to the
lack of statistics. Second, in the case of quadratic coupling 
one should take into account the screening
effect
when relating the bounds on $\Lambda_2$
from binary pulsars to the parameters in the particle
physics Lagrangian. Indeed, a quadratically coupled scalar field acquires
effective mass $m^2_{\rm eff}\sim \rho/\Lambda_2^2$ inside an object
with density $\rho$. If the corresponding Compton wavelength is
shorter than the size of the object, the field inside the object gets
frozen at $\Phi=0$ and only an outer layer of width $m_{\rm eff}^{-1}$
interacts with ULDM. This can degrade the bounds on $\Lambda_2^{-1}$
by a few orders of magnitude compared to those shown on
Fig.~\ref{fig:cons2} \cite{longpap}. 

A peculiarity of the binary pulsar constraints is that every single
system is sensitive to ULDM masses only in a few narrow bands
corresponding to resonances on the first $N\lesssim 10$
harmonics.  Conservatively one can require that the system stays in resonance during the whole observational campaign, so that the changes in $P_b$ induced by ULDM accumulate over the time; this would maximize the sensitivity of observations to the effect.  This yields an  estimate of the band
width   
$\delta \omega \sim 5\times 10^{-23}\, \mathrm{eV}/$(years of observation).
While this is much smaller than the total mass range of interest, 
the $O(10^3)$ binary systems with different periods expected to be
discovered by SKA \cite{Kramer:2015bea} will allow for a significant
coverage.

\paragraph*{Acknowledgments} We thank Vitor Cardoso,  Emmanuel Fonseca, Paulo Freire, Gian Giudice,
Michelangelo Mangano, 
David Pirtskhalava
and Valery Rubakov for discussions. The work of S.S. is
supported by the Swiss National Science Foundation.

\bibliographystyle{hieeetr}
\bibliography{biblio}

\begin{thebibliography}{10}

\bibitem{Marsh:2015xka}
D.~J.~E. Marsh, ``{Axion Cosmology},'' {\em Phys. Rept.}, vol.~643, pp.~1--79,
  2016, 1510.07633.

\bibitem{Peccei:1977hh}
R.~D. Peccei and H.~R. Quinn, ``{CP Conservation in the Presence of
  Instantons},'' {\em Phys. Rev. Lett.}, vol.~38, pp.~1440--1443, 1977.

\bibitem{Wilczek:1977pj}
F.~Wilczek, ``{Problem of Strong p and t Invariance in the Presence of
  Instantons},'' {\em Phys. Rev. Lett.}, vol.~40, pp.~279--282, 1978.

\bibitem{Weinberg:1977ma}
S.~Weinberg, ``{A New Light Boson?},'' {\em Phys. Rev. Lett.}, vol.~40,
  pp.~223--226, 1978.

\bibitem{Graham:2015cka}
P.~W. Graham, D.~E. Kaplan, and S.~Rajendran, ``{Cosmological Relaxation of the
  Electroweak Scale},'' {\em Phys. Rev. Lett.}, vol.~115, no.~22, p.~221801,
  2015, 1504.07551.

\bibitem{Svrcek:2006yi}
P.~Svrcek and E.~Witten, ``{Axions In String Theory},'' {\em JHEP}, vol.~06,
  p.~051, 2006, hep-th/0605206.

\bibitem{Arvanitaki:2009fg}
A.~Arvanitaki, S.~Dimopoulos, S.~Dubovsky, N.~Kaloper, and J.~March-Russell,
  ``{String Axiverse},'' {\em Phys.Rev.}, vol.~D81, p.~123530, 2010, 0905.4720.

\bibitem{Arvanitaki:2014faa}
A.~Arvanitaki, J.~Huang, and K.~Van~Tilburg, ``{Searching for dilaton dark
  matter with atomic clocks},'' {\em Phys. Rev.}, vol.~D91, no.~1, p.~015015,
  2015, 1405.2925.

\bibitem{Hlozek:2014lca}
R.~Hlo\v{z}ek, D.~Grin, D.~J.~E. Marsh, and P.~G. Ferreira, ``{A search for
  ultralight axions using precision cosmological data},'' {\em Phys. Rev.},
  vol.~D91, no.~10, p.~103512, 2015, 1410.2896.

\bibitem{Hlozek:2016lzm}
R.~Hlo\v{z}ek, D.~J.~E. Marsh, D.~Grin, R.~Allison, J.~Dunkley, and
  E.~Calabrese, ``{Future CMB tests of dark matter: Ultralight axions and
  massive neutrinos},'' {\em Phys. Rev.}, vol.~D95, no.~12, p.~123511, 2017,
  1607.08208.

\bibitem{Hu:2000ke}
W.~Hu, R.~Barkana, and A.~Gruzinov, ``{Cold and fuzzy dark matter},'' {\em
  Phys.Rev.Lett.}, vol.~85, pp.~1158--1161, 2000, astro-ph/0003365.

\bibitem{Amendola:2005ad}
L.~Amendola and R.~Barbieri, ``{Dark matter from an ultra-light
  pseudo-Goldsone-boson},'' {\em Phys. Lett.}, vol.~B642, pp.~192--196, 2006,
  hep-ph/0509257.

\bibitem{Bozek:2014uqa}
B.~Bozek, D.~J.~E. Marsh, J.~Silk, and R.~F.~G. Wyse, ``{Galaxy UV-luminosity
  function and reionization constraints on axion dark matter},'' {\em Mon. Not.
  Roy. Astron. Soc.}, vol.~450, no.~1, pp.~209--222, 2015, 1409.3544.

\bibitem{Schive:2015kza}
H.-Y. Schive, T.~Chiueh, T.~Broadhurst, and K.-W. Huang, ``{Contrasting Galaxy
  Formation from Quantum Wave Dark Matter, $\psi$DM, with $\Lambda$CDM, using
  Planck and Hubble Data},'' {\em Astrophys. J.}, vol.~818, no.~1, p.~89, 2016,
  1508.04621.

\bibitem{Sarkar:2015dib}
A.~Sarkar, R.~Mondal, S.~Das, S.~Sethi, S.~Bharadwaj, and D.~J.~E. Marsh,
  ``{The effects of the small-scale DM power on the cosmological neutral
  hydrogen (HI) distribution at high redshifts},'' {\em JCAP}, vol.~1604,
  no.~04, p.~012, 2016, 1512.03325.

\bibitem{Hui:2016ltb}
L.~Hui, J.~P. Ostriker, S.~Tremaine, and E.~Witten, ``{Ultralight scalars as
  cosmological dark matter},'' {\em Phys. Rev.}, vol.~D95, no.~4, p.~043541,
  2017, 1610.08297.

\bibitem{Khmelnitsky:2013lxt}
A.~Khmelnitsky and V.~Rubakov, ``{Pulsar timing signal from ultralight scalar
  dark matter},'' {\em JCAP}, vol.~1402, p.~019, 2014, 1309.5888.

\bibitem{Porayko:2014rfa}
N.~K. Porayko and K.~A. Postnov, ``{Constraints on ultralight scalar dark
  matter from pulsar timing},'' {\em Phys. Rev.}, vol.~D90, no.~6, p.~062008,
  2014, 1408.4670.

\bibitem{Marsh:2015daa}
D.~J.~E. Marsh, ``{Nonlinear hydrodynamics of axion dark matter: Relative
  velocity effects and quantum forces},'' {\em Phys. Rev.}, vol.~D91, no.~12,
  p.~123520, 2015, 1504.00308.

\bibitem{Arvanitaki:2010sy}
A.~Arvanitaki and S.~Dubovsky, ``{Exploring the String Axiverse with Precision
  Black Hole Physics},'' {\em Phys.Rev.}, vol.~D83, p.~044026, 2011, 1004.3558.

\bibitem{Arvanitaki:2016qwi}
A.~Arvanitaki, M.~Baryakhtar, S.~Dimopoulos, S.~Dubovsky, and R.~Lasenby,
  ``{Black Hole Mergers and the QCD Axion at Advanced LIGO},'' {\em Phys.
  Rev.}, vol.~D95, no.~4, p.~043001, 2017, 1604.03958.

\bibitem{Aoki:2016kwl}
A.~Aoki and J.~Soda, ``{Detecting ultralight axion dark matter wind with laser
  interferometers},'' {\em Int. J. Mod. Phys.}, vol.~D26, p.~1750063, 2017,
  1608.05933.

\bibitem{Derevianko:2013oaa}
A.~Derevianko and M.~Pospelov, ``{Hunting for topological dark matter with
  atomic clocks},'' {\em Nature Phys.}, vol.~10, p.~933, 2014, 1311.1244.

\bibitem{VanTilburg:2015oza}
K.~Van~Tilburg, N.~Leefer, L.~Bougas, and D.~Budker, ``{Search for ultralight
  scalar dark matter with atomic spectroscopy},'' {\em Phys. Rev. Lett.},
  vol.~115, no.~1, p.~011802, 2015, 1503.06886.

\bibitem{Stadnik:2015kia}
Y.~V. Stadnik and V.~V. Flambaum, ``{Can dark matter induce cosmological
  evolution of the fundamental constants of Nature?},'' {\em Phys. Rev. Lett.},
  vol.~115, no.~20, p.~201301, 2015, 1503.08540.

\bibitem{Hees:2016gop}
A.~Hees, J.~Gu\'ena, M.~Abgrall, S.~Bize, and P.~Wolf, ``{Searching for an
  oscillating massive scalar field as a dark matter candidate using atomic
  hyperfine frequency comparisons},'' {\em Phys. Rev. Lett.}, vol.~117, no.~6,
  p.~061301, 2016, 1604.08514.

\bibitem{Stadnik:2016zkf}
Y.~V. Stadnik and V.~V. Flambaum, ``{Improved limits on interactions of
  low-mass spin-0 dark matter from atomic clock spectroscopy},'' {\em Phys.
  Rev.}, vol.~A94, no.~2, p.~022111, 2016, 1605.04028.

\bibitem{Graham:2015ifn}
P.~W. Graham, D.~E. Kaplan, J.~Mardon, S.~Rajendran, and W.~A. Terrano, ``{Dark
  Matter Direct Detection with Accelerometers},'' {\em Phys. Rev.}, vol.~D93,
  no.~7, p.~075029, 2016, 1512.06165.

\bibitem{Arvanitaki:2015iga}
A.~Arvanitaki, S.~Dimopoulos, and K.~Van~Tilburg, ``{Sound of Dark Matter:
  Searching for Light Scalars with Resonant-Mass Detectors},'' {\em Phys. Rev.
  Lett.}, vol.~116, no.~3, p.~031102, 2016, 1508.01798.

\bibitem{Stadnik:2014tta}
Y.~V. Stadnik and V.~V. Flambaum, ``{Searching for dark matter and variation of
  fundamental constants with laser and maser interferometry},'' {\em Phys. Rev.
  Lett.}, vol.~114, p.~161301, 2015, 1412.7801.

\bibitem{Stadnik:2015xbn}
Y.~V. Stadnik and V.~V. Flambaum, ``{Enhanced effects of variation of the
  fundamental constants in laser interferometers and application to dark matter
  detection},'' {\em Phys. Rev.}, vol.~A93, no.~6, p.~063630, 2016, 1511.00447.

\bibitem{Geraci:2016fva}
A.~A. Geraci and A.~Derevianko, ``{Sensitivity of atom interferometry to
  ultralight scalar field dark matter},'' {\em Phys. Rev. Lett.}, vol.~117,
  no.~26, p.~261301, 2016, 1605.04048.

\bibitem{Arvanitaki:2016fyj}
A.~Arvanitaki, P.~W. Graham, J.~M. Hogan, S.~Rajendran, and K.~Van~Tilburg,
  ``{Search for light scalar dark matter with atomic gravitational wave
  detectors},'' 2016, 1606.04541.

\bibitem{Manchester:2015mda}
R.~N. Manchester, ``{Pulsars and Gravity},'' {\em Int. J. Mod. Phys.},
  vol.~D24, no.~06, p.~1530018, 2015, 1502.05474.

\bibitem{Kramer:2016kwa}
M.~Kramer, ``{Pulsars as probes of gravity and fundamental physics},'' {\em
  Int. J. Mod. Phys.}, vol.~D25, no.~14, p.~1630029, 2016, 1606.03843.

\bibitem{Wex:2014nva}
N.~Wex, ``{Testing Relativistic Gravity with Radio Pulsars},'' 2014, 1402.5594.

\bibitem{Will:2014xja}
C.~M. Will, ``{The Confrontation between General Relativity and Experiment},''
  {\em Living Rev.Rel.}, vol.~17, p.~4, 2014, 1403.7377.

\bibitem{Hui:2012yp}
L.~Hui, S.~T. McWilliams, and I.-S. Yang, ``{Binary systems as resonance
  detectors for gravitational waves},'' {\em Phys. Rev.}, vol.~D87, no.~8,
  p.~084009, 2013, 1212.2623.

\bibitem{Bertotti1973}
B.~{Bertotti}, ``{Is the Solar System Gravitationally Closed?},'' {\em
  Astrophys. Lett.}, vol.~14, p.~51, 1973.

\bibitem{1975SvA....19..270R}
V.~N. {Rudenko}, ``{Test bodies under the effect of gravitational radiation},''
  {\em Sov. Astron.}, vol.~19, p.~270, Apr. 1975.

\bibitem{1978ApJ...223..285M}
B.~{Mashhoon}, ``{On tidal resonance},'' {\em Astrophys. J.}, vol.~223,
  pp.~285--298, July 1978.

\bibitem{Turner:1979yn}
M.~S. Turner, ``{Influence of a weak gravitational wave on a bound system of
  two point masses},'' {\em Astrophys. J.}, vol.~233, pp.~685--693, 1979.

\bibitem{Mashhoon:1981wn}
B.~Mashhoon, B.~J. Carr, and B.~L. Hu, ``{The Influence of Cosmological
  Gravitational Waves on a Newtonian Binary System},'' {\em Astrophys. J.},
  vol.~246, pp.~569--591, 1981.

\bibitem{Pani:2015qhr}
P.~Pani, ``{Binary pulsars as dark-matter probes},'' {\em Phys. Rev.},
  vol.~D92, no.~12, p.~123530, 2015, 1512.01236.

\bibitem{pscat}
{\it ATNF Pulsar Catalogue},
  \url{http://www.atnf.csiro.au/people/pulsar/psrcat/}.

\bibitem{2005AJ....129.1993M}
R.~N. {Manchester}, G.~B. {Hobbs}, A.~{Teoh}, and M.~{Hobbs}, ``{The Australia
  Telescope National Facility Pulsar Catalogue},'' {\em AJ}, vol.~129,
  pp.~1993--2006, Apr. 2005, astro-ph/0412641.

\bibitem{Kramer:2015bea}
M.~Kramer and B.~Stappers, ``{Pulsar Science with the SKA},'' 2015, 1507.04423.

\bibitem{Kehl:2016mgp}
M.~S. Kehl, N.~Wex, M.~Kramer, and K.~Liu, ``{Future measurements of the
  Lense-Thirring effect in the Double Pulsar},'' in {\em Proceedings, 14th
  Marcel Grossmann Meeting on Recent Developments in Theoretical and
  Experimental General Relativity, Astrophysics, and Relativistic Field
  Theories (MG14) (In 4 Volumes): Rome, Italy, July 12-18, 2015}, vol.~2,
  pp.~1860--1865, 2017, 1605.00408.

\bibitem{Damour:1991rd}
T.~Damour and J.~H. Taylor, ``{Strong field tests of relativistic gravity and
  binary pulsars},'' {\em Phys. Rev.}, vol.~D45, pp.~1840--1868, 1992.

\bibitem{longpap}
D.~Blas, D.~L\'opez~Nacir, and S.~Sibiryakov.
\newblock {\it to appear}.

\bibitem{Navarro:1995iw}
J.~F. Navarro, C.~S. Frenk, and S.~D.~M. White, ``{The Structure of cold dark
  matter halos},'' {\em Astrophys. J.}, vol.~462, pp.~563--575, 1996,
  astro-ph/9508025.

\bibitem{Schive:2014dra}
H.-Y. Schive, T.~Chiueh, and T.~Broadhurst, ``{Cosmic Structure as the Quantum
  Interference of a Coherent Dark Wave},'' {\em Nature Phys.}, vol.~10,
  pp.~496--499, 2014, 1406.6586.

\bibitem{Bertotti:2003rm}
B.~Bertotti, L.~Iess, and P.~Tortora, ``{A test of general relativity using
  radio links with the Cassini spacecraft},'' {\em Nature}, vol.~425, p.~374,
  2003.

\bibitem{Olive:2007aj}
K.~A. Olive and M.~Pospelov, ``{Environmental dependence of masses and coupling
  constants},'' {\em Phys. Rev.}, vol.~D77, p.~043524, 2008, 0709.3825.

\bibitem{Madsen:2012rs}
E.~C. Madsen, I.~H. Stairs, M.~Kramer, F.~Camilo, G.~B. Hobbs, G.~H. Janssen,
  A.~G. Lyne, R.~N. Manchester, A.~Possenti, and B.~W. Stappers, ``{Timing the
  main-sequence-star binary pulsar J1740-3052},'' {\em Mon. Not. Roy. Astron.
  Soc.}, vol.~425, p.~2378, 2012, 1207.2202.

\bibitem{Freire:2010tf}
P.~C.~C. Freire {\em et~al.}, ``{On the nature and evolution of the unique
  binary pulsar J1903+0327},'' {\em Mon. Not. Roy. Astron. Soc.}, vol.~412,
  p.~2763, 2011, 1011.5809.

\bibitem{Freire:2007jd}
P.~C.~C. Freire, S.~M. Ransom, S.~Begin, I.~H. Stairs, J.~W.~T. Hessels, L.~H.
  Frey, and F.~Camilo, ``{Eight New Millisecond Pulsars in NGC 6440 and NGC
  6441},'' {\em Astrophys. J.}, vol.~675, p.~670, 2008, 0711.0925.

\bibitem{Lynch:2011aa}
R.~S. Lynch, P.~C.~C. Freire, S.~M. Ransom, and B.~A. Jacoby, ``{The Timing of
  Nine Globular Cluster Pulsars},'' {\em Astrophys. J.}, vol.~745, p.~109,
  2012, 1112.2612.

\bibitem{Janssen:2008mh}
G.~H. Janssen, B.~W. Stappers, M.~Kramer, D.~J. Nice, A.~Jessner, I.~Cognard,
  and M.~B. Purver, ``{Multi-telescope timing of PSR J1518+4904},'' {\em
  Astron. Astrophys.}, vol.~490, p.~753, 2008, 0808.2292.

\bibitem{Freire:2007xg}
P.~C.~C. Freire, A.~Wolszczan, M.~v.~d. Berg, and J.~W.~T. Hessels, ``{A
  Massive Neutron Star in the Globular Cluster M5},'' {\em Astrophys. J.},
  vol.~679, p.~1433, 2008, 0712.3826.

\bibitem{Fonseca:2014qla}
E.~Fonseca, I.~H. Stairs, and S.~E. Thorsett, ``{A Comprehensive Study of
  Relativistic Gravity using PSR B1534+12},'' {\em Astrophys. J.}, vol.~787,
  p.~82, 2014, 1402.4836.

\bibitem{Weisberg:2016jye}
J.~M. Weisberg and Y.~Huang, ``{Relativistic Measurements from Timing the
  Binary Pulsar PSR B1913+16},'' {\em Astrophys. J.}, vol.~829, no.~1, p.~55,
  2016, 1606.02744.

\bibitem{Ferdman:2014rna}
R.~D. Ferdman {\em et~al.}, ``{PSR J1756-2251: a pulsar with a low-mass neutron
  star companion},'' {\em Mon. Not. Roy. Astron. Soc.}, vol.~443, no.~3,
  pp.~2183--2196, 2014, 1406.5507.

\bibitem{Bhat:2008ck}
N.~D.~R. Bhat, M.~Bailes, and J.~P.~W. Verbiest, ``{Gravitational-radiation
  losses from the pulsar-white-dwarf binary PSR J1141-6545},'' {\em Phys.
  Rev.}, vol.~D77, p.~124017, 2008, 0804.0956.

\bibitem{vanLeeuwen:2014sca}
J.~van Leeuwen {\em et~al.}, ``{The Binary Companion of Young, Relativistic
  Pulsar J1906+0746},'' {\em Astrophys. J.}, vol.~798, no.~2, p.~118, 2015,
  1411.1518.

\bibitem{Kramer:2006nb}
M.~Kramer {\em et~al.}, ``{Tests of general relativity from timing the double
  pulsar},'' {\em Science}, vol.~314, pp.~97--102, 2006, astro-ph/0609417.

\bibitem{2003ApJ...599..806A}
J.~W. {Armstrong}, L.~{Iess}, P.~{Tortora}, and B.~{Bertotti}, ``{Stochastic
  Gravitational Wave Background: Upper Limits in the 10$^{-6}$ to 10$^{-3}$ Hz
  Band},'' {\em \apj}, vol.~599, pp.~806--813, Dec. 2003.

\end{thebibliography}
\end{document}